# An Exciton-Polariton Fano Resonance Driven by Second Harmonic Generation

Yafeng Wang[1], Liming Liao[1], Tao Hu[1], Song Luo[1], Lin Wu[1], Jun Wang[1], Zhe Zhang[1], Wei Xie[1], Liaoxin Sun[1], A.V.Kavokin[2,3,4], Xuechu Shen[1], Zhanghai Chen[1*]

[1]*State Key Laboratory of Surface Physics, Key Laboratory of Micro and Nano Photonic Structures (Ministry of Education), Department of Physics, Fudan University, Shanghai 200433, China and Collaborative Innovation Center of Advanced Microstructures, Nanjing University, Najing Jiangsu 210093, China*

[2]*Physics and Astronomy School, University of Southampton, Highfield, Southampton, SO249QH, United Kingdom*

[3]*SPIN-CNR, Viale del Politechnico 1, I-00133, Rome, Italy*

[4]*Spin Optics Laboratory, St-Petersburg State University, 1 Ulianovskaya, St-Petersburg, 198504, Russia*

Angle-resolved second harmonic generation (SHG) spectra of ZnO microwires show characteristic Fano resonances in the spectral vicinity of exciton-polariton modes. The output SHG spectra after SHG interacting with exciton polariton shows a resonant enhancement peak accompanied by a suppression dip originating from the constructive and destructive interference respectively. It is demonstrated that the Fano line shape, and thus the Fano asymmetry parameter q, can be tuned by the phase-shift of the two channels. The phase-dependent q was calculated and the model describes our experimental results well. In particular, the phase-to-q relation unveil the crucial information about the dynamics of the system, e.g., defining the line shape of output SHG spectra in a superposition of quantum states.

Fano resonance, known as a fingerprint of quantum interference effect, has attracted much attention since discovered by Ugo Fano.[1] It originates from the interference between a discrete state and a continuum of quantum states, giving rise to a characteristic asymmetric emission line shape. First introduced to describe the atomic photo-ionization [1, 2], exploration of this important phenomenon and the underlying physics have been extended to the fields of photonic crystals [3, 4], plasmonic devices[5-7], metamaterials[8, 9], fiber-cavity systems [10], Raman scattering[11, 12], nonlinear optics, etc. [13-20]. The Fano line shape is considered as a unique asymmetric spectral response, and its tunability in a discrete-continuum coupled quantum system provides great opportunities for developing optronic devices such as on chip optical sensors and switching devices [15, 20, 21]. A crucial parameter that characterizes a Fano resonance is the spectroscopic asymmetry factor q. The spectral reflectance and transmittance at resonance of a discrete-continuum coupled quantum system depend strongly on this factor. Recently, a universal approach for manipulating the asymmetry parameter q of a Fano line shape, i.e., phase shift tuning of the discrete state, has been introduced by Christian Ott *et al.* in an atomic system [17, 18]. It is expected that in an optical resonator with discrete optical modes, this approach of phase shift tuning can be easily employed by varying the shape, size, etc. of resonators.

In microcavities with semiconductor gain media, well-defined cavity-exciton polariton modes dominate the optical properties near the band-edge energy. As half-light-half-matter quasiparticles, exciton polaritons are formed due to the strong coupling of photons and excitons. These quasiparticles have a bosonic nature and a tunable effective mass. In the scenario of cavity quantum electrodynamics, polariton modes in a semiconductor microcavity not only provide a playground for simulating cold atom physics, such as Bose-Einstein condensation, superfluidity, etc. in the solid state environment at elevated temperatures [22-25], but also can serve as discrete states for Fano resonance studies. However, despite its great importance for optoelectronics, the Fano resonance in a cavity polariton system is yet to be demonstrated. The main challenge is to search for coherent broad continuum states that can efficiently couple with a cavity polariton state. Nonlinear optical effects, such as second harmonic generation (SHG), may provide a channel for preparation of coherent continuum states. Efforts have been made to study the polariton-SHG interaction in GaAs based microcavities, and two-photon absorption of polaritons using SHG excitation has been observed [26]. However, no sign of Fano resonances has been reported so far.

Here we report on the observation of Fano resonances in SHG spectra of ZnO microwires. As a wide band-gap semiconductor, ZnO has not only large exciton binding energy (~60 meV at room temperature (RT)), but also large nonlinear coefficients, providing both stable cavity exciton-polariton states and strong SHG continuum at RT. In the past few years, these two aspects have attracted much attentions. Striking features of cavity polariton effects, such as ultra-strong exciton-photon coupling [23], polariton lasing at RT and above RT (25, 27-29), polariton parametric scattering [27], polariton-polariton coupling [30], polariton-phonon interaction [31, 32], and evaporative cooling of polaritons [33] etc. have been revealed. On the other hand, SHG effect of ZnO has also been well explored, including the exciton enhanced SHG [34], SHG is influenced by crystalline structure, dimensions, grains and doping [35], second harmonic polaritons [36] etc. Taking advantage of coherent polariton states and SHG continuum states with comparable strength in ZnO, Fano resonances with high coupling coefficients can be expected. In a whispering gallery modes (WGM) resonators of ZnO, the gain medium acts also as the optical cavity, which provides a unique system for the large overlap in real space of a discrete cavity polariton mode and broad SHG continuum. We reveal the Fano resonances between polariton modes and SHG. The effect is manifested in the appearance of a narrow peak accompanied by a suppression dip originating from the constructive and destructive interference between the two emission channels, respectively. The angle-resolved spectroscopy enables us to observe the clear Fano fingerprint. Symmetric and asymmetric line shapes are obtained due to the phase shift of the polariton mode in the coherent coupling. The phase-dependent q has been extracted from these data. The phase-to-q relation unveiled the crucial information about the dynamics of the system [2].

Fig. 1a shows the scanning electron microscope (SEM) of a typical ZnO microwire. The diameter of the ZnO microwire is about 1.6 μm and the c-axis is parallel to the longitudinal crystal axis. Fig. 1b shows a schematic configuration of our experiment. A ZnO microwire was laid on a glass substrate, the fundamental laser beam was focused on the back and SHG was detected from the front side. Angle-resolved spectra measurements were performed along the angles θ and ϕ by using a Princeton Instrument SP 750 spectrometer with the wavelength resolution of 0.02 nm and a charge-coupled detector (CCD) of Pixis 100. The detection scheme is shown in Scheme S1 in the Supplemental material (SM). For the SHG measurements, femtosecond laser pulses from a Ti: sapphire laser (150 fs) served as the fundamental light source. The repetition rate is 80 MHz and the wavelength can be tuned from 700 nm to 1000 nm. All the experiments were performed at room temperature.

Fig. 1c shows the energy dispersion of 29[th] and 28[th] TM (E // c-axis) polariton modes under continuous He-Cd laser (325 nm) excitation. The polariton modes show a narrow line width and can be fitted with a Lorentzian function (shown in Fig. S1 in SM) [38]. The N=29 polariton mode has a quality factor of ~2430, which is obtained from the ratio $E/\Delta E$, where $\Delta E \approx 1.3$ meV is the full width at half maximum (FWHM). The pure SHG radiation without coupling with a polariton mode is displayed in Fig. S2a and S2b in the SM. SHG spectra measured with the $\bar{\text{polarization}}$ resolution is shown in Fig. S2c. The line shape of the SHG also can be well fitted by using a Lorentzian function, giving the FWHM≈12 meV. It should be emphasized that the FWHM of SHG is about 9.2 times larger than that of the polariton mode, therefore it makes background SHG playing role of the continuum states and the polariton mode being a discrete state.

In our experiments, the power of the fundamental wave is set to 3.4 mJ/cm$^2$ in order to minimize the PL emission that may interfere with the SHG radiation. The wavelength of SHG was set to resonate with a polariton mode. The resulting spectra are shown in Fig. 2a. Both TE and TM polariton modes were taken into account, as the energy of the 28$^{th}$ TE and 29$^{th}$ TM modes are quite close: 3.157 eV and 3.169 eV, respectively (see Fig. S5 in SM) [38], while only the TM polariton mode was found to be populated due to the polarization selection rules. When the broad SHG band scans from the lower energy side to the high energy side near 3.169 eV appears a strong peak accompanied by a dip adjacent to it. This is a signature of a Fano resonance. The corresponding spectra at θ=0° are shown in Fig. 2b. Their asymmetric line shapes signify that SHG and the polariton mode are not independent in ZnO. Instead, they interfere with each other to form a new hybrid emission in the vicinity of the resonance. In this regime, the polariton and SHG not only are highly coherent but also have a large overlap in real space which results in large coupling efficiency.

Here we need to mention that, in the absence of the polariton mode, SHG has a broad symmetric line shape with its intensity independent on the angle ϕ (Shown in Fig. S3 in SM). However, ones SHG and polariton are in the regime of Fano resonance, the asymmetric line shape varies with the angle ϕ. As shown by the angle ϕ resolved spectra of SHG resonated with the 29$^{th}$ polariton mode in Fig. 3a (the SHG-polariton mode resonance spectra for the 28$^{th}$ and 27$^{th}$ modes are shown in Fig. 4 SM). The pattern shows that the spectral intensity distribution depends on the detection angle ϕ. As the ϕ changes from -30° to 30°, the positions of the dip (dark) and the peak (bright) swap. This phenomenon indicates that the interference between the SHG and polariton mode controls the line shape of the spectra. To show distinctly the line shape variation with varying the angle ϕ, we plot the spectral profiles at some typical angles (-27°, -15°, -8.3°, 0°, 7.7°, 15°, 27°) in Fig. 3c (black dotted curves). The phenomenological line shape under Fano resonance condition is given by:

$$I = A_F[\frac{(\varepsilon+q)^2}{\varepsilon^2+1}-1]+I_b \tag{1}$$

where $I$ is the emission spectral intensity. $A_F[\frac{(\varepsilon+q)^2}{\varepsilon^2+1}-1]$ is the Fano resonance term [1]. $A_F$ is the amplitude coefficient. $\varepsilon= 2(\omega-\omega_r)/\Gamma_p$ is the reduced energy, where $\omega_r$ is the resonant frequency and $\Gamma_p$ is the linewidth of the polariton mode. q denotes the Fano asymmetry factor. The second term $I_b$ is the background offset of the broad continuum states contributed by SHG. The line profiles are fitted well using equation (1) (red solid curves) and the q values are obtained for each curve. According to Fano, the dimensionless parameter q introduced in equation (1) manifests the ratio of the transition probabilities to the polariton state and to the continuous SHG states, which can be either positive or negative depending on the phase difference of the two transition channels. It is this parameter q that governs the resonant line shape. In Fig 3c, the spectra at -15°, 0° and 15° show characteristic asymmetric line shapes, in these cases the probabilities for these two transition channels are similar, leading to q near ±1. The spectra at angles -8.3° and 7.7° show a nearly symmetric line shape with a dip at the center, which is known as anti-resonance or EIT-like effect in other systems [7]. This is because the value of q approaches zero here, in which case the optical transition to the continuum SHG states dominates. When it comes to -27° and 27°, the line shapes show a peak at the resonant position, due to the large absolute q values of -3.8±0.3 and -4.2±0.4. This q variation is a direct consequence of the phase-difference changes in the interference between the two channels, the above results indicate that the relative phase-difference, and hence the degree of asymmetry of the resonant line shape depends on the angle ϕ. In order to extract the phase-difference dependence of the polariton-SHG Fano resonance, we plot the q values for each ϕ in Fig. 3b. It is noticeable that q is positive from -24.7° to -8.3° (and 7.7° to 25.5°), and negative in the other area. Remarkably, q shows a symmetric dependence on the angle ϕ. As far as we know, the SHG background generated through such a ZnO microwire can be considered as light radiating from a hexagonal medium, which has a phase that hardly changes with angle ϕ. Therefore we attribute the variation of q with the angle ϕ to the phase shift of the polariton channel.

To describe the phase of the polariton mode quantitatively, we calculate the near field electric field distribution for the TM polarized 29$^{th}$ polariton mode emission in Fig. 4a using the finite element analysis (FEA) method [35]. The x and y are real space coordinates. In our simulation, the polariton effect manifests itself in the wavelength dependence of the effective refractive index [23]. One can see in Fig. 4a that the electric field (including its phase) of the polariton mode has a strong dependence on the angle ϕ. The inset in Fig. 4a shows the equiphase surface (the black curves) for the area in the white rectangle. The phase is zero for the dashed lines and π for the solid lines. In order to find out precisely the angle corresponding to the boundary separating opposite phases in the momentum space, we applied the Fourier transform for Fig. 4a and obtained the far field emission pattern in Fig. 4b. The horizontal and vertical coordinates are wavevector projections $\mathbf{k}_x$ and $\mathbf{k}_y$, respectively. The emission pattern shows a circle consisted of many bright maxima characterized by the identical $|\mathbf{k}| = (\sqrt{\mathbf{k}_x^2 + \mathbf{k}_y^2})$. The gaps between the adjacent bright spots indicate the boundaries where the phase change sign. Using Fig. 4a and Fig. 4b, we have obtained the dependence of the phase φ on the angle ϕ shown in Fig. 4c. The marked angles, consistent with those in Fig. 4b, indicate that the phase stays above π/2 from -7.5° to 7.5°, then it changes almost to -π and remains slightly below -π/2 from 7.5° to 23° (and -7.5° to -23°), finally it goes sharply to π/2 beyond 23° or -23°.

To obtain more insight on the coupling features mentioned above, here we model the experiment in more details. The time-dependent electric field can be expressed as:

$$E(t) = A_P e^{-\frac{\Gamma_P}{2}t} e^{-i\omega_P t + i\varphi} + A_S e^{-\frac{\Gamma_S}{2}t} e^{-i\omega_S t}, \qquad (2)$$

where $A_P$ and $A_S$ are emission field amplitude of the polariton state and the SHG state. $\Gamma_p$ and $\Gamma_s$ are the decay rates of the two channels, respectively. $\omega_p$ and $\omega_s$ are frequencies of the two channels. φ is the phase shift of the polariton state. In the frequency domain, the frequency-dependent electric field intensity $I_\omega$ is obtained via the Fourier transformation of Equation (2), which yields

$$I_\omega = |E(\omega)|^2 \propto \left[ \frac{A_P^2}{\Gamma_P^2(1+\varepsilon_P^2)} + \frac{A_S^2}{\Gamma_S^2(1+\varepsilon_S^2)} \right.$$

$$+2\frac{A_P A_S}{\Gamma_P \Gamma_S}\frac{(1+\varepsilon_P\varepsilon_S)\cos\varphi+(\varepsilon_P-\varepsilon_S)\sin\varphi}{(1+\varepsilon_P^2)(1+\varepsilon_S^2)}\Bigg] \quad (3)$$

Here $\varepsilon_p=(\omega-\omega_p)/\Gamma_p$ and $\varepsilon_s=(\omega-\omega_s)/\Gamma_s$. The first and the second terms in the square bracket describe the symmetric Lorentzian shape in the frequency domain originating from the polariton mode and the SHG, respectively, and the last term describes the coupling strength between these two characters. Using this equation, we simulated the emission pattern as shown in Fig. 4d, and the effective q is obtained by fitting every line profile in Fig. 4e (red solid curve) using equation (1). These theoretical results show good agreement with the experimental results shown in Fig. 3b. Thereafter for the sake of a simplicity, $\Gamma_s$ is approximated to infinity, which means the SHG is replaced by an infinitely broad background state, and the amplitude of the background state is much larger than that of the discrete state ($A_s/\Gamma_s \gg A_p/\Gamma_p$), which provides a strong enough continuum of states. In this limit, the relation between q and the phase shift $\varphi$ can be obtained as [35]:

$$q = -\cot\frac{\varphi}{2} \quad (4)$$

This result means that by manipulating the phase $\varphi$ of the discrete state one can directly alter the asymmetry parameter q, which is similar to the results obtained in an atomic system [17]. Knowing this, with use of the dependence of the phase $\varphi$ on the angle $\phi$ in Fig. 4c, we mapped the angle $\phi$ into the q parameter and plotted q as a function $\phi$ in Fig. 4e (green dashed curve). It agrees with the experimental values of q well. This indicates that Eq. (4) is sufficiently accurate in our system. Having analytical expression, we are able to understand the variation of the complex SHG line shape with the angle $\phi$. The phase $\varphi$-sensitive Fano resonance exhibits a rich physics of the polariton-SHG coupling.

In conclusion, we have experimentally demonstrated and theoretically analyzed a Fano resonance in the SHG radiation spectra of an individual semiconductor microwire. In the coupling between exciton-polariton and SHG, the polariton mode serves as a discrete state and the SHG serves as the continuum. Angle-resolved spectra exhibit pronounced variation of the Fano line shape, resulting from the phase shift of the cavity polariton modes. These observations are important for the precision control of the Fano lineshapes in semiconductor microwires.

The work is funded by National Science Foundation for China (No. 11225419, No. 91321311 and No. 11474297) and Program of Shanghai Subject Chief Scientist (No. 14XD1400200). AK thanks the HORIZON 2020 RISE project CoExAn (Grant No. 644076) and the Russian Foundation for Basic Research, grant 15-59-30406-PT.

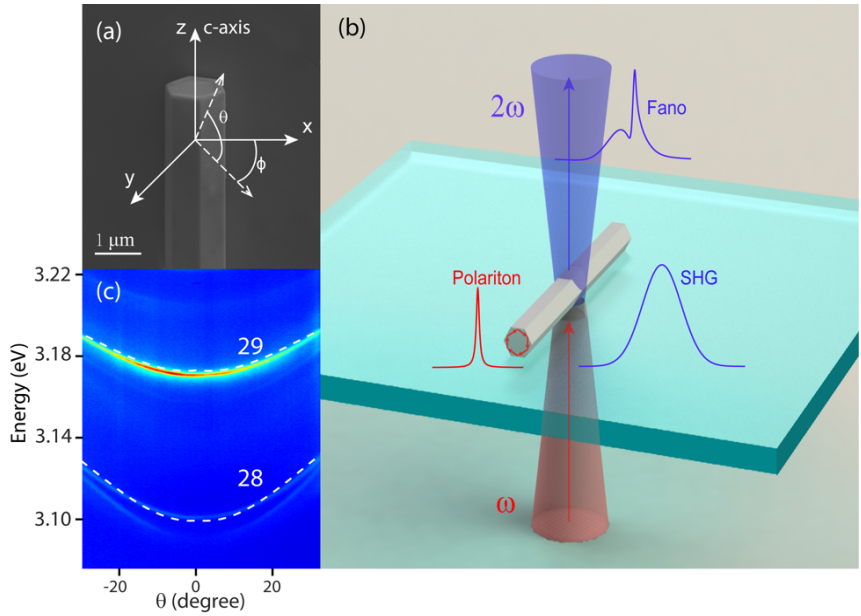

**FIG. 1**. (a) The scanning electronic image (SEM) of a typical ZnO microwire. The angles defined as θ and φ represent the two detection planes of the angle resolved spectra. (b) The fundamental light wave irradiates the ZnO microwire from back of the glass, the SHG radiation is emitted from the top of the wire. (c) The dispersion of the 28$^{th}$ and 29$^{th}$ TM polarized polariton.

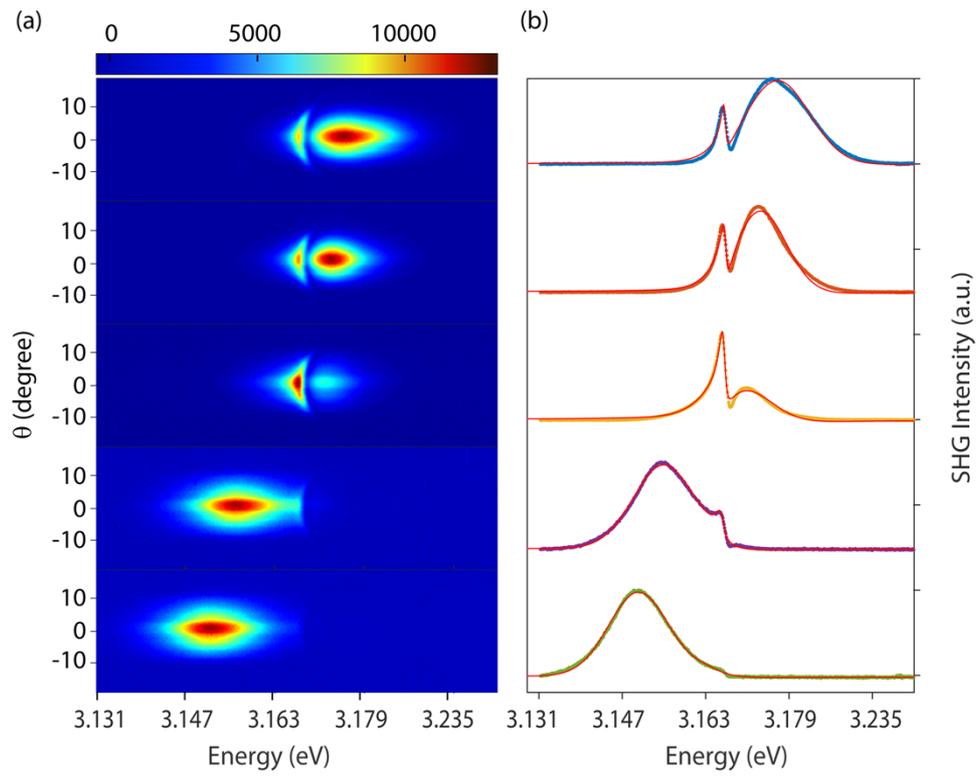

**FIG. 2**. (a): SHG spectra with the peak at 3.175 eV, 3.172 eV, 3.169 eV, 3.155 eV, 3.151 eV from up to down. (b): the corresponding line curves taken out at θ= 0°.

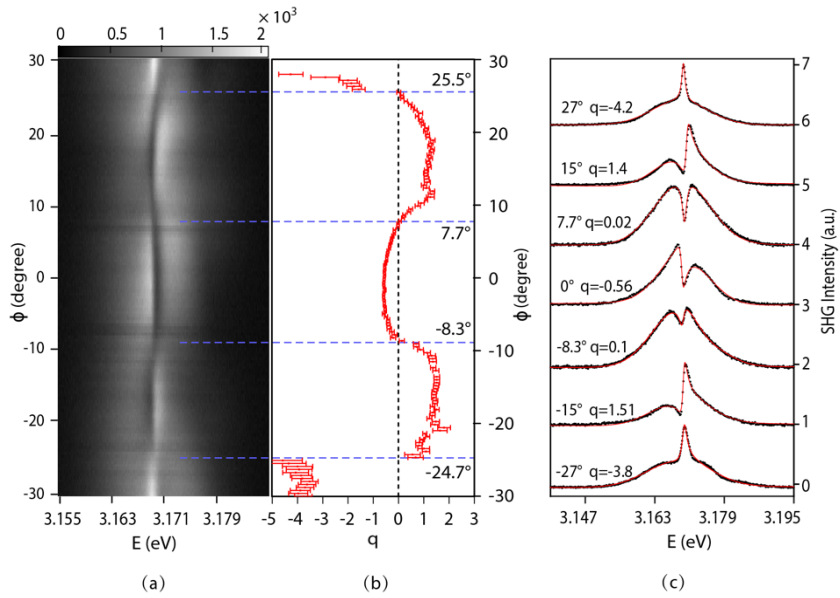

**FIG. 3**. (a): The angle ϕ resolved spectral pattern. SHG and the polariton mode are resonant at 3.169 eV. (b): q values at each angle are obtained by fitting each line profile. The dashed lines are the boundaries where q changes its sign. (c) The typical line profiles taken at ϕ=-27°, -15°, -8.3°, 0°, 7.7°, 15°, 27° from pattern (a). The black dotted lines are experimental data and the red solid lines are fitted data. These curves show the different line shapes resulting from the Fano coupling in the system.

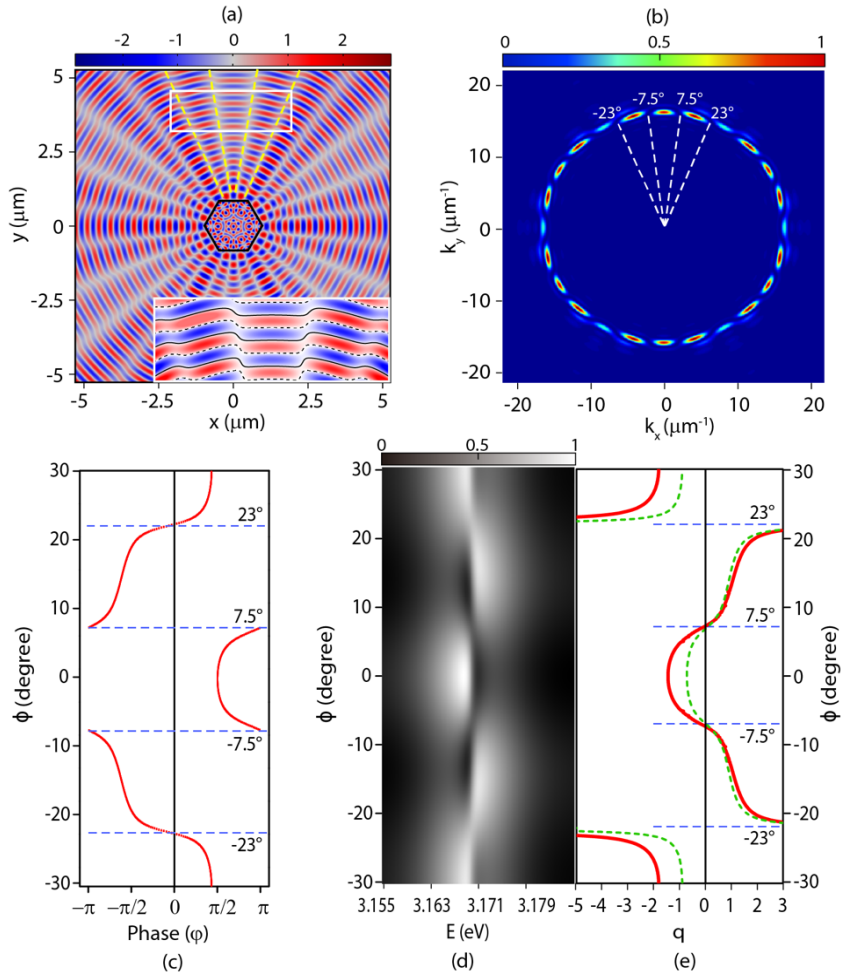

**FIG. 4** (a): The electric field distribution pattern for the 29$^{th}$ polariton mode in the real space calculated using finite element analysis. Inset: the equiphase surface marked by the dashed and solid lines for the white rectangle area. (b): The Fourier Transform of pattern (a) for the whole space. The dashed lines marked the angles where the phase of the emission electric field in pattern (a) change sign. (c): The phase φ v.s. ϕ curve obtained through the phase information in (a). (d): The theoretical pattern calculated using equation (3) with $A_p/\Gamma_p$ being the amplitude against ϕ obtained from (b), and $A_s/\Gamma_s$ being the same order of $A_p/\Gamma_p$. $\Gamma_p$ and $\Gamma_s$ are the FWHM (in frequency domain) of the 29$^{th}$ polariton mode (1.3 meV) and the SHG (12 meV) respectively. (e): The q variation obtained by fitting each line of pattern d (red solid line) and q(φ)=-cot(φ/2) (blue dashed line).